\documentclass[11pt]{article}

\newfont{\goth}{eufm10 scaled\magstep1}
\newfont{\tenmsb}{msbm10 scaled\magstep1}

\let\ssection=\section
\renewcommand{\section}{\setcounter{equation}{0}\ssection}

\def\br{{\bf r}}

\def\D{D\llap{\big/}}
\def\pib{{\bf\pi}}
\def\sigmab{{\bf\sigma}}

\def\rot{{\bf rot\ }}
\def\2{{\frac{1}{2}}}

\usepackage{graphicx}

\begin{document}

\setlength{\baselineskip}{16pt}

\title{Helicity-supersymmetry of dyons}

\author{
F. BLOORE\\
DAMTP, The University of Liverpool\\
P. O. Box
147, LIVERPOOL L69 3BX (U.K.)
\\
[4pt]
P.~A.~Horv\'athy
\\
Laboratoire de Math\'ematiques et de Physique Th\'eorique
\\ 
Universit\'e de Tours.
\\
Parc de Grandmont. 
F-37 200 TOURS (France)\footnote{e-mail~:horvathy@univ-tours.fr}.
}

\date{\today}

\maketitle

\begin{abstract}
The 'dyon' system of D'Hoker and Vinet consisting of a spin $\2$
particle with anomalous gyromagnetic ratio $4$ in the combined field of a Dirac
monopole plus a Coulomb plus a suitable $1/r^2$ potential (which
arises in the long-range limit of a self-dual monopole) is studied
following Biedenharn's approach to the Dirac-Coulomb problem: the
explicit solution is obtained using the `Biedenharn-Temple operator', 
$\Gamma$, and  the extra 
two-fold degeneracy is explained by the subtle supersymmetry generated by
the 'Dyon Helicity' or generalized `Biedenharn-Johnson-Lippmann' operator 
${\cal R}$. 
The new SUSY anticommutes with the chiral SUSY discussed previously.
\end{abstract}

Tours Preprint $N^o$ 34/91.  September 1991.
{\sl Journ. Math. Phys.} {\bf 33}, 1869 (1992)

\noindent

\section{Introduction}

In a recent series of papers [1] D'Hoker and Vinet studied the
strange 'dyon' system consisting of a
charged, spin $\2$ particle with anomalous gyromagnetic ratio 4 in a combined
Dirac monopole + Coulomb potential + inverse-square potential field, 
described by the Hamiltonian
\begin{equation}
H_1={\pib}^2+q^2(1-{1\over r})^2- 
2q{\sigmab\cdot\br\over r^3} .
\end{equation}
Here 
$
\pib = -i{\bf \partial} -e{\bf A}$, 
\ ${\bf A}$ being the vector-potential
of a Dirac monopole of strength $g$, $\rot \,{\bf A} = -g{\bf r}/r^3$, 
$q = eg > 0$ without loss of generality. The
surprising dynamical and supersymmetries
allow to calculate the spectrum [1] (and the $S$-matrix [2]), 
\begin{equation} 
E_p=q^2\Big(1-{q^2\over p^2}\Big), \  
\qquad 
p=q, q+1, \ldots 
\end{equation}
whose multiplicity is $2(p^2 - q^2)$ for $p \geq q+1$, and $2q$ for $p = q$.

The
Hamiltonian $H_1$ is similar to that of the 'MIC-Zwanziger' [3] system,    
\begin{equation}
H_0={\bf\pib}^2+q^2(1-{1\over r})^2 ,
\end{equation}
representing a spin $0$ particle in the same field. 
Doubling  $H_0$ yields  a spin $\2$ particle but with gyromagnetic ratio $g=0$,
which has the same symmetries, spectrum and multiplicities as (1.1) (except
for the $0$-energy ground states: for MIC-Zwanziger, $p \geq q+1$ only) [4]. 

This is  explanained by that $H_1$ and $H_0$ are {\it
superpartners}  [1,4] : they both arise as the long-distance
limits of a Dirac particle in the field of a self-dual monopole which
has a {\it chiral supersymmetry},  
\begin{equation}
\left(\begin{array}{ll}
H_1 &\\  &H_0
\end{array}\right)
=\D^2 
\quad
\hbox{for} \quad
\D=\left(\begin{array}{ll} &Q^{\dagger}\\ Q &
\end{array}\right)
 =
\left(\begin{array}{ll} &{\bf\sigmab.\pib}-i\Phi\\ {\bf\sigmab.\pib}+i\Phi
\end{array}\right),
\end{equation}
where $\Phi=q(1-1/r)$ is the long-range tail of the Higgs field.  
The 'upper' and 'lower' sectors (i.e. the $\pm1$ eigensectors of
the chirality operator
$\rho_3$) are related by the unitary transformations
\begin{equation}
U=Q^{\dagger}\ {1\over\sqrt{H_0}}
\qquad  
U^{-1}=U^{\dagger} = 
{1\over\sqrt{H_0}}\ Q,
\end{equation}
which intertwine $H_1=Q^{\dagger} Q$ and $H_0=QQ^{\dagger}$, 
$H_1=UH_0U^{\dagger}$. If $\Psi_0$ is an
$H_0$-eigenfunction with eigenvalue $E>0$, then
\begin{equation}
\left(\begin{array}{c}
U\Psi_0\\
\pm\Psi_0\\
\end{array}
\right)
\end{equation}
is a $\D$-eigenfunction with eigenvalues $\pm \sqrt {E}$. The arousal of
zero-energy ground states for $H_1$ (but not for $H_0$) as solutions of 
$Q\Psi = 0$ [4] is a
nice manifestation of supersymmetry. The multiplicity $2q$ is
consistent with the index-theoretical calculations in a self-dual
monopole background [5]. 

Although this approach provides an insight into the
mysteries found by D'Hoker and Vinet, explicit calculations are rather
complicated, because $U$ mixes the radial and
angular parts.
In another paper [6] we introduced two operators,
\begin{equation}
x={\bf\sigmab}.{\bf L}+1,
\qquad
y={\bf\sigmab}.{\bf L}+1+2q{\bf\sigmab}.{\hat r},
\end{equation}
where ${\bf L}={\bf\ell}-q{\hat{\bf r}}$,  
${\bf\ell}={\bf r}\times{\bf\pib}$ is the orbital angular
momentum. $y$ (resp. $x$) are constants of the motion for the $H_1$ (resp $H_0$)
dynamics, and satisfy  
\begin{equation}
y^2=x^2={\bf J}^2+1/4. 
\end{equation}
Both $y$ and $x$ have therefore
eigenvalues $\pm (j+ 1/2)$ and bring $H_1$ and $H_0$ into a
{\it non-relativistic Coulomb} form (cf. Section 3),  
\begin{equation}
-(\partial_r+{1\over r})^2-{2q^2\over r}+q^2
+ 
{1\over r^2}
\left(\begin{array}{ll}(j-{1\over2})(j+{1\over2})&\\
&(j+{1\over2})(j+{3\over2})
\end{array}\right),
\end{equation}
whose spectrum is
shown on FIGs. 1 and 2.
\begin{figure}
\includegraphics{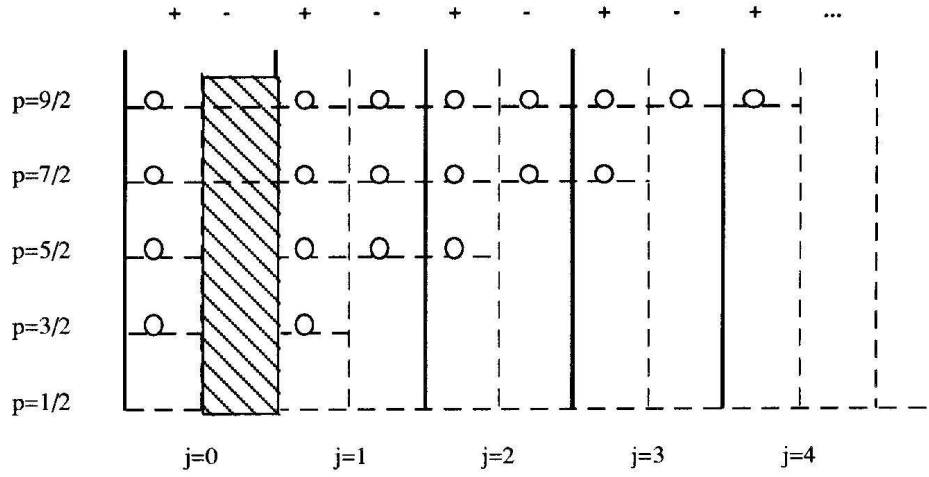}
\label{Bloore1}
\caption{\it The dyon spectrum in the $g=0$ sector. The sign refers to that
of $(-x)$. Each $j\geq q+1/2$ sector is doubly degenerate. For $j=q-1/2$
there are no $(-x)=-q$ states. The energy only depends on the principal
quantum number $ =L(\gamma)+1+n$.}
\end{figure}
\begin{figure}
\includegraphics{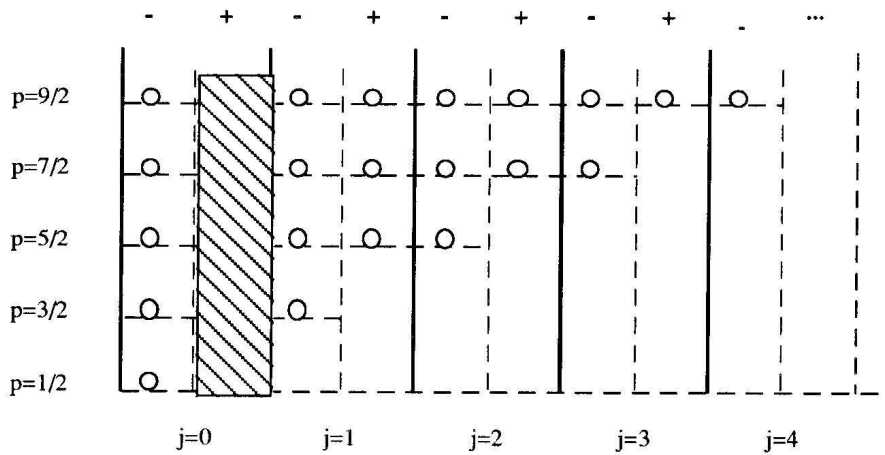}
\label{Bloore2}
\caption{\it The dyon spectrum in the $g=4$ sector. The sign refers to that 
of $(-y)$. Each $j \geq q + 1/2$ sector is doubly degenerate. For $j=q-1/2$
there are no $(-y) = +q$ states but $E=0$ ground states arise for 
$(-y)=-q$.}
\end{figure}

It is tempting
to think that 'supersymmetry' means just $x \mapsto -y$. However,
$U^{\dagger}xU \ne -y$ which seemingly contradicts our expectations. But we
show below that our system admits {\it another supersymmetry} 
encountered
before in Biedenharn's approach to the Dirac-Coulomb problem [7-9]. This
second SUSY is actually more convenient then the chiral one, because it does
respect the angular decomposition.

The method is based on two operators, namely the
'Biedenharn-Temple' operator $\Gamma$ [7, 10] and the
'Biedenharn-Johnson-Lippmann' [11] or 'Coulomb Helicity' [7-9] operator ${\cal
R}$. The first of these allows us to write the iterated Dirac equation in a
non-relativistic Coulomb form, but with an {\it irrational angular
momentum $\ell(\gamma)$}
(where $\gamma$ is an eigenvalue of the Biedenharn-Temple operator), yielding the
well-known spectrum. The irrationality of $\ell(\gamma)$ results in shifting
the different angular momentum sectors with respect to each other i.e. in the
fine structure.  The two-fold degeneracy of each $j=const$. sector, (except for
the ground state) can be viewed as indicating SUSY, the intertwining
transformation being Biedenharn's 'Coulomb Helicity operator' ${\cal
R}$ [8, 12]. 

In this paper we extend the Biedenharn method to the dyon problem. 
Remarkably, things work even better than in
the Dirac-Coulomb case: one gets
again a Coulomb-type equation, but with {\it (half)integer} angular momentum
$L(\gamma)$:  The fine structure is automatically suppressed, without having to
be removed by hand, as Biedenharn did for his `symmetric Hamiltonian' [7].

We show first that it  reproduces the `$x-y$' picture of Ref. [4].
Then we focus our attention to the $j = const.$ sectors of FIG. 1-2.
The
pattern is reminiscent to that of a supersymmetric system - except
that the ground states have non-zero energy $E_0^{(j)}$ cf. (4.1). 
But this can
easily be cured: it is enough to subtract $E_0^{(j)}$ from 
the Hamiltonian and consider rather the new Hamiltonians $K_1$ and $K_0$
defined by 
\begin{equation}
\left(\begin{array}{ll}K_1&\\&K_0
\end{array}\right)
=\D^2-E_0^{(j)}
=\left(\begin{array}{ll}
H_1-q^2+{q^4 \over(j+1/2)^2}&\\ 
&H_0-q^2+{q^4\over(j+1/2)^2}
\end{array}\right).
\end{equation}

In Section 4 we show that this indeed yields a SUSY system~:
following Biedenharn, we exhibit a new conserved operator,
\begin{equation}
{\cal S}=\left(\begin{array}{ll} 
&S^{\dagger}\\ S&\end{array}\right) 
=
\left(\begin{array}{ll} \displaystyle
&i{\sigmab.\pib}+{q\over\displaystyle r}+({\sigmab}.{\hat {\bf
r}}) {q^2 \over\displaystyle x}
\\
i{\sigmab.\pib}-{q\over\displaystyle r}+({\bf\sigmab}. 
{\hat {\bf r}}){q^2\over\displaystyle y}&
\end{array}\right)
\end{equation}
which we call 'Dyon Helicity', and show that it is  a  
supersymmetry operator for the new Hamiltonian,
$$ 
{\cal S}^2=\left(\begin{array}{ll}K_1 &\\&K_0\end{array}\right).
$$ 

The need for an extra supersymmetry is
understood by noting that giving the energy, the total angular momentum
and the third component, $E$, $j$,  and $\mu$, respectively, do not specify completely a state: one
has to give also the sign of $\gamma$ -- and this is exactly this sign which
labels the new SUSY sectors.

Our Dyon Helicity operator ${\cal S}$ also allows us to derive, along the
lines indicated in Refs. [8, 9], the $S$-matrix (Section 5). Thing again work
better as for the Dirac-Coulomb problem, where an arbitrary phase has to
be chosen in each $j = const.$ sector [8,9]. Here, since the dyon quantum
numbers are half-integers rather then irrational, it is enough to chose a phase
in one single sector.

Another simplification with respect to the Dirac-Coulomb problem is that we
work with the mass $0$ Dirac equation -- albeit in higher dimension -- and
mass enters only after dimensional reduction.

Let us mention that the same technique applies to a particle in the field
of a charged monopole [13-17]. The difference with dyons comes from
that the Coulomb potential belongs to the time
coordinate with Minkowski signature, in constrast as for  dyons, where the Higgs
field appears in an extra euclidean dimension. This changes a sign and
quantities simply add, yielding perfect squares rather than irrational (or even
complex) values.

\section {The Biedenharn-Temple operator for dyons.}

Let us consider a massless Dirac particle in the long-distance field 
${\bf B}=-q{\bf r}/r^3$, $\Phi=q(1-1/r)$ of a
Bogomolny-Prasad-Sommerfield monopole. Identifying $\Phi$ with the fourth
component of a gauge field we get a self-dual Yang-Mills
field in four dimensions. This leads to the Dirac Hamiltonian\footnote{here
$\rho_1 =\left(\begin{array}{ll} &1_2\\ 1_2 &\end{array}\right)$, 
$\rho_2 =\left(\begin{array}{ll} &-i1_2\\ i1_2 &\end{array}\right)$, 
$\rho_3=\left(\begin{array}{ll}1_2 &\\&-1_2\end{array}\right)$.}  
\begin{equation}
\D=\rho_1({\sigmab\cdot\pib})-\rho_2\Phi\, 
=
\left(\begin{array}{ll}
 &Q^{\dagger}
 \\ Q &\end{array}\right)
=
\left(\begin{array}{ll}&{\sigmab.\pib}-i\Phi\\
{\sigmab.\pib}+i\Phi&
\end{array}\right).
\end{equation}
Unlike in the
Coulomb case, the scalar term $\rho_2 \Phi$ is now {\it off-diagonal}, because
it comes from the fourth, euclidean, direction rather then from the time
coordinate.

The total angular momentum, 
\begin{equation}
{\bf J}={\bf L}+{{\bf\sigmab}\over 2},\qquad 
{\bf L}={\bf \ell}-q {\hat{\bf r}}, 
\qquad {\bf \ell}={\bf r}\times{\bf\pib}
\end{equation}
is conserved.
Set\footnote{Notice that z is ${\bf\sigmab.L}+1+qw$ [13], rather then 
$\sigmab\cdot L+1$, as in the Dirac case [17].}  
\begin{equation}
w={\bf\sigmab}.{\hat{\bf r}}, 
\qquad
z={\bf\sigmab.\ell}+1 
\qquad {\cal K}=-\rho_2z=-\rho_2({\bf\sigmab.\ell}+1).
\end{equation}
Note that $w^2=1$ and that $z$
anticommutes with $w$ and ${\sigmab.\pib}$,  
\begin{equation}
\lbrace z, w\rbrace=0
\qquad
\lbrace z, 
{\bf\sigmab.\pib}\rbrace=0.
\end{equation}
Since $z$ anticommutes with the first term in
eqn. (2.1) and commutes with the second, ${\cal K}$ commutes  with the
Dirac Hamiltonian $\D$. Using 
$$
(\sigmab\cdot L)^2={\bf L}^2+i
\sigmab({\bf L}\times{\bf L})={\bf L}^2-\sigmab\cdot L
$$ 
one proves that 
\begin{equation}
{\cal K}^2=z^2={\bf J}^2+{1\over4}-q^2,
\end{equation}
so that z (and $\cal K$) have {\it irrational}{}\footnote{In the Dirac-Coulomb case
the eigenvalues of ${\cal K}$ are half-integers.} eigenvalues,  
\begin{equation}
\kappa=\sqrt{(j+1/2)^2-q^2} \ , 
\end{equation} 
(${\bf J}^2 = j(j+1)$). Since $j \geq q -
1/2$, ${\cal K}$ is hermitian, but for $j = q -1/2$ its eigenvalue $\kappa$
vanishes and thus ${\cal K}$ is not invertible.
 
The Dirac operator $(2.1)$ is, as in any even dimensional space,
chiral-supersymmetric: for $Q$ and $Q^{\dagger}$ in (1.4), $\lbrace
Q, Q^{\dagger}\rbrace$ is a SUSY Hamiltonian and the SUSY sectors are the $\pm1$
eigenspaces of the chirality operator $\rho_3 $. The supercharges $Q$ and
$Q^{\dagger}$ can be written as 
\begin{equation} 
Q=-iw\Big(\partial_r+{1\over r}-{z+qw\over r}+qw
\Big)=-i\Big(\partial_r+{1\over r}+{z-qw\over r}+qw\Big) w,
\end{equation}
and
\begin{equation}
Q^{\dagger} = iw \Big(-(\partial_r + {1\over r}) + {z - qw \over r} + qw \Big) =
i \Big(- (\partial_r + {1\over r}) - {z + qw \over r} + qw \Big)w .
\end{equation}
The square $\D^2$ of the Dirac Hamiltonian (2.1) is thus
\begin{equation}
\left(\begin{array}{ll}Q^{\dagger} Q & 
\\
 &Q Q^{\dagger}\end{array}\right)
=
 - (\partial _r + {1 \over r})^2 - {2q^2 \over r} + q^2 + 
{z^2 + q^2 \over r^2} - {1 \over r^2} 
\left(\begin{array}{ll}z + qw & \\
 &z - qw\end{array}\right).
\end{equation}

Let us now introduce the {\it Biedenharn-Temple operator}  
\begin{equation}
\Gamma = - (z + \rho_3 qw) = - ({\sigmab . \ell} + 1 + \rho_3 qw).
\end{equation}
Although $\Gamma$ does not commute with the Dirac Hamiltonian $\D$, it
commutes with its square $\D^2$: it is thus conserved for the quadratic
dynamics $H_0$ and $H_1$ but not for the Dirac Hamiltonian.
In terms of $\Gamma$, $\D^2$ becomes 
\begin{equation}
\D^2 = - (\partial _r + {1 \over r})^2+ 
{\Gamma(\Gamma+1)\over r^2} - {2q^2\over r}+q^2.
\end{equation}
Eqns. (2.5) and (2.10) imply
\begin{equation}
\Gamma^2 = z^2 + q^2 = {\bf J}^2 + {1\over4},
\end{equation}
so that the eigenvalues of $\Gamma$ are {\it (half)integers}, 
\begin{equation}
\gamma = \pm(j+1/2) \qquad \hbox{sign}\ \gamma = \hbox{sign}\ \kappa .
\end{equation}
Hence for a $\Gamma$-eigenfunction (constructed in the next Section),
\begin{equation}
\Gamma (\Gamma + 1) = L(\gamma) (L(\gamma) + 1) \quad \hbox{where} \quad
L(\gamma) = j \pm {1 \over 2} \ ,
\end{equation}
(The sign is plus or minus depending on the sign of $\gamma$). Observe
that $L(\gamma)$ is now a (half)integer\footnote{For Dirac-Coulomb
 the eigenvalues of $\Gamma$ are irrational, yielding the
fine structure.}. Using the notations  
$ x = z - qw$ and $y = z + qw$, cf. (1.7), the supercharges are written as 
\begin{equation}
Q = -iw \Big(\partial_r + {1\over r} - {y \over r} + qw \Big)=  
-i \Big( \partial_r + {1\over r} + {x \over r} + qw \Big) w 
\end{equation}
and
\begin{equation}
Q^{\dagger} = iw \Big(-(\partial_r + {1\over r}) + {x \over r} + qw \Big) =
i \Big(-(\partial_r + {1\over r}) - {y \over r} + qw \Big) .
\end{equation}
One can also write
\begin{equation}
\Gamma  =
 -\left(\begin{array}{ll}
  \sigmab \cdot L + 1 + 2qw &0 
 \\ 0 &\sigmab\cdot L + 1
 \end{array}\right)
= \left(\begin{array}{ll}
- y &0 \\ 0 &-x
 \end{array}\right). 
\end{equation} 
$x$ and $y$ are self-adjoint,  
$x = x^{\dagger}$, $y = y^{\dagger}$, $w = w^{\dagger}$. In terms of $x$ and
$y$ the lower (resp. upper) two components of (2.9) are exactly the
Hamiltonians $H_0$ and $H_1$ in [4]. 

\section {An explicit solution.}

In order to find an explicit solution, we first construct angular 2-spinors
$\varphi_{\pm}^{\mu}$ and  $\Phi_{\pm}^{\mu}$, which are both eigenfunctions of 
${\bf J}^2$ and $J_3$ with eigenvalues $j(j+1)$ and $\mu$, and which diagonalize
the operators $x$ and $y$:
\begin{equation}
x \varphi_{\pm}^{\mu} = \mp\mid\gamma\mid \varphi^{\mu}_{\pm}
\qquad
y \Phi_{\pm}^{\mu} = \mp\mid\gamma\mid \Phi^{\mu}_{\pm}.
\end{equation}

In the lower ($\rho_3 = -1$) sector the gyromagnetic ratio is $g = 0$, so the
$x$ - eigenspinors $\varphi$ are obtained by the prescription valid in
the Coulomb case except for that the ordinary spherical harmonics should be
replaced by the Wu and Yang [19] 'monopole' harmonics. The coefficient of the
$r^{-2}$ term here is the square of the orbital angular momentum, 
\begin{equation} 
x(x-1) = {\bf L}^2 =
L(\gamma)(L(\gamma)+1), 
\end{equation}
so that $L(\gamma)$ is just the orbital angular quantum number.
Due to the addition theorem of the angular momentum, if $j \geqÊ
q+1/2$, $L(\gamma) = j \pm 1/2$ but for $j= q-1/2$
the only allowed value of $L(\gamma)$ is $L(\gamma) = j+1/2$.

In detail, for $j \geq q + 1/2$ consider therefore the spinorial functions 
\begin{equation}
\varphi_{\pm}^{\mu} = \sqrt{{L(\gamma) + 1/2 \pm \mu \over 2L(\gamma) + 1}}\, 
Y^{\mu-1/2}_{L(\gamma)}
	\left(\begin{array}{l}1 \\ 0
	\end{array}\right) 
\pm 
\sqrt{{L(\gamma) + 1/2 \mp \mu \over 2L(\gamma) + 1}}\, 
Y^{\mu +1/2}_{L({\gamma})}
	\left(\begin{array}{l}0 \\ 1\end{array}\right) ,
\end{equation} 
where the $Y$'s are monopole spherical harmonics and the
sign $\pm$ refers to the sign of $\gamma$. The $\varphi$'s 
satisfy\footnote{the superscript $\mu$ is dropped for the sake of simplicity.} 
$$
{\bf J}^2 \varphi_{\pm} = j(j+1) \, 
\varphi_{\pm}
\qquad
J_3 \varphi_{\pm} = \mu \varphi_{\pm},
\quad
\mu = -j, \cdots, j,
\qquad
 {\bf L}^2 \varphi_{\pm} = L(\gamma)(L(\gamma)+1)\varphi_{\pm}.
$$
Since ${\bf L . \sigmab} = {\bf J}^2 - {\bf L}^2 - 3/4$, 
$$
x \ \varphi_{\pm} = \Big( {\bf L . \sigmab} + 1\Big) \ \varphi_{\pm} = \mp
\mid \gamma \mid \varphi_{\pm},
$$ 
as wanted.

For $j=q-1/2$ no $\varphi_-$ (i.e. $L(\gamma) = q-1$) state is available but
eqn. (3.3) still yields $(2q)$ $\varphi_+$-states with $L(\gamma) = q$:
\begin{equation}
\varphi_+^{\mu}
=\sqrt{{q + 1/2 \pm \mu \over 2q + 1}}\, Y^{\mu-1/2}_q
\left(\begin{array}{l}1\\ 0\end{array}\right) 
\pm 
\sqrt{{q + 1/2 \mp \mu \over 2q + 1}}\, 
Y^{\mu +1/2}_q
	\left(\begin{array}{l}0 \\ 1\end{array}\right)
\end{equation} 
(where $\mu=-(q-1/2),\ldots,(q-1/2)$) are eigenstates of $x$ with
eigenvalue $(-q)$.

The $y$-eigenspinors $\Phi$ of the upper (i.e. $\rho_3 = 1$) sector are
constructed indirectly. Assume first that one can find angular spinors
$\chi_{\pm}$ which diagonalize 
$z={\bf\sigmab}\cdot{\bf\ell} + 1$,  
\begin{equation}
z\ \chi_{\pm}=\pm\mid\kappa\mid
\chi_{\pm},
\end{equation}
and also satisfy 
${\bf J}^2\chi_{\pm}^{\mu}=j(j+1) \, \chi_{\pm}^{\mu}$, 
$j=q-1/2, q+1/2,\cdots$, 
$J_3\chi_{\pm}^{\mu}=\mu\ \chi_{\pm}^{\mu}$, \ $\mu = -j, \cdots, j$ and
\begin{equation}
w\ \chi_{\pm}^{\mu}=\chi_{\mp}^{\mu}.
\end{equation}

In the subspace spanned by the $\chi_{\pm}$'s, $x=z-qw$ and 
$y=z+qw$ have the remarkably symmetric matrix representations
\begin{equation}
\Big[x\Big]=\left(\begin{array}{ll}\mid\kappa\mid &-q \\ -q &-\mid\kappa\mid
\end{array}\right)
\qquad
\Big[y\Big]=\left(\begin{array}{ll}\mid\kappa\mid &q \\ q &-\mid\kappa\mid
\end{array}\right).
\end{equation}
The eigenvectors $\varphi_{\pm}$ and $\Phi_{\pm}$ of $x$ and $y$ with
eigenvalues $\pm \mid \gamma \mid$ are thus
\begin{equation}
\begin{array}{lll}
&\varphi_+=(\mid\kappa\mid+\mid\gamma\mid)\chi_+-q\chi_- 
&\varphi_-=\ q\chi_++(\mid\kappa\mid+\mid\gamma\mid)\chi_-
\\ & & \\
&\Phi_+=(\mid\kappa\mid+\mid\gamma\mid)\chi_++q\chi_- 
&\Phi_-=-q\chi_++(\mid\kappa\mid+\mid\gamma\mid)\chi_-
\end{array}.
\end{equation}
Expressing the $\chi$'s from the upper two equations in terms of the
$x$-eigenspinors $\varphi$ yield the z-eigenspinors
\begin{equation}
\chi_+={1\over2\mid\gamma\mid}\Big(\varphi_+ 
+{q\over{\mid\gamma\mid+\mid\kappa\mid}}
\ \varphi_-\Big)
\qquad
\chi_-={1\over2\mid\gamma\mid}\Big(- 
{q\over{\mid\gamma\mid+\mid\kappa\mid}} 
\ \varphi_++
\varphi_-\Big),  
\end{equation}
which {\it do} indeed satisfy (3.5). For $j=q-1/2$, $\chi_-$ is missing and
$\chi_+$ is proportional to the lowest $\varphi_+$ as expressed in (3.4), 
since no $\varphi_-$ is available.

Eliminating the $\chi'$s allows to deduce the
$y$-eigenspinors $\Phi$ from the $x$-eigenspinors $\varphi$ according to  
\begin{equation}
\Phi_+={1\over\mid\gamma\mid}
\Big(\mid\kappa\mid\varphi_++q\varphi_-\Big)
\qquad
\Phi_-={1\over\mid\gamma\mid}
\Big(-q\varphi_++\mid\kappa\mid\varphi_-\Big) 
\end{equation}
which (by construction) satisfy 
${\bf J}^2\Phi_{\pm}=j(j+1) 
\Phi_{\pm}$,\ $J_3\Phi_{\pm}=\mu\Phi_{\pm},\,
\mu=-j,\ldots, j$ and
$$
y\ \Phi_{\pm}=\mp\mid\gamma\mid\Phi_{\pm}.
$$
Finally, $w = {\sigmab}.{\hat {\bf r}}$ interchanges
the $x$ and $y$ eigenspinors, 
\begin{equation}
w\,\varphi_{\pm}^{\mu}=\Phi^{\mu}_{\mp}.
\end{equation}

In contrast to what happens in the `lower' (i.e. $\rho_3=-1$) sector,
in the `upper (i.e. $\rho_3=1$) sector 
$y(y-1)={\bf L}^2-2{\bf\sigmab\cdot{\hat r}}$ 
is {\it not} the square of an angular momentum 
and hence we
{\it do} have $L(\gamma)=q-1$ states:  $\mid\gamma\mid=q$, 
$\kappa=0$
for the lowest value of total angular momentum, $j=q-1/2$, and for $\gamma =
-q$ eqn. (3.8) yields (3.4), 
\begin{equation} 
\Phi_0\, \, (=\Phi_-)=\varphi_+,
\end{equation}
while the entire $\Phi_+$ -tower is missing.
By (3.6), this 
is a $(-1)$ eigenstate of $w$,
\begin{equation}
w \ \Phi_0=-\Phi_0.
\end{equation}
Since $\varphi_+$ is a $(-q)$ eigenstate of $x$, 
$\Phi_0$ is indeed an
eigenstate of $y=x+2qw$ with eigenvalue $(+q)$. Since
\begin{equation}
\Gamma (\Gamma +1)\ \Phi^{\mu}_{\gamma}= 
L(\gamma)(L(\gamma) + 1)\ \Phi^{\mu}_{\gamma}\ , 
\qquad
\Gamma (\Gamma +1)\ \varphi^{\mu}_{\gamma}= 
L(\gamma)(L(\gamma)+1)\ \varphi^{\mu}_{\gamma},
\end{equation} 
by construction, for $j \geq q+1/2$ the eigenfunctions of $\D^2$ are found as 
\begin{equation}
\left.\begin{array}{lll}
\Psi_{\pm\mid\gamma\mid} &= u_{\pm} 
\left(\begin{array}{l}\Phi_{\pm}\\ 0 \end{array}\right)
\quad
&\hbox{for} \ \ \rho_3 = 1, 
\\[8pt]
\psi_{\pm\mid\gamma\mid} &= u_{\pm} 
\left(\begin{array}{l}0 \\ \varphi_{\pm}\end{array}\right)   
&\hbox{for} \ \ \rho_3 = -1 
\end{array}\right\} 
\qquad \hbox{if} \quad j \geq q+1/2, 
\end{equation}
where the radial functions $u_{\pm}(r)$ solve the non-relativistic
Coulomb-type equations 
\begin{equation}
\Big[- (\partial _r + {1 \over r})^2 + {L(\gamma)(L(\gamma) +1) \over r^2}
 - {2q^2 \over r} + q^2 \Big] \ u_{\pm} = E^2u_{\pm}.  
\end{equation}
By (2.14), these are just the upper (resp. lower)
equations in (1.9), and hence
\begin{equation}
u_{\pm}(r) \, \propto \, r^{L(\gamma)}e^{ikr} \
F \Big( L(\gamma)+1-i{q^2 \over k}, 2L(\gamma) + 2, -2ikr \Big),
\end{equation}
where $k = \sqrt{E^2 - q^2}$. For $j = q-1/2$ we get the $(2q)$ spinors  
\begin{equation}
\psi_+ = u_+\left(\begin{array}{ll}
0 \\ \varphi_+
\end{array}\right), \qquad \hbox{sign} \ \gamma = +1
\end{equation}
in the $\rho_3 = -1$ sector with $L(\gamma) = q$\footnote{$L(\gamma) = q$-values
arise in the $\rho_3 = 1$ sector for $\gamma = - (q + 1)$.}, with $u_+$ 
still as in (3.17). 

The energy levels are obtained from the poles of $F$,
$L(\gamma) + 1 - iq^2/k =  -n$, $n = 0, 1, \ldots$. Introducing the principal
quantum number $p = L(\gamma) + 1 + n \geq q+1$ we conclude that, in both
$\rho_3$ sectors,    
\begin{equation}
E_p = q^2 \Big(1 - ({q\over p})^2 \Big), \qquad p =  \ q+1,...
\end{equation}
The same energy is obtained if $L + n = L' + n'$. The degeneracy of a
$p \geq q+1$-level is hence $2 (p^2 - q^2)$. 

If $j=q-1/2$, $(2q)$ extra states arise in the $\rho_3 = 1$ sector for $\gamma
= -q$, 
\begin{equation}
\Psi_0 = u_0 
\left(\begin{array}{ll}\Phi_{0} \\ 0
\end{array}\right) 
\qquad \hbox{for} \ \ \rho_3 = 1 
\quad \hbox{and} \quad
\gamma = -q ,
\end{equation}
where $u_0$ solves (3.16) with $L(\gamma) = q-1$.
The principal quantum number is now $p = q$  yielding the $2q$-fold
degenerate $0$ - energy ground states. Since $F(0, a, z) = 1$, and the lowest
wave number $k_0$ is $iq$, $u_0$ is simply
\begin{equation}
u_0 = r^{q-1} e^{-q r} ,
\end{equation}
cf. [1, 4]. The situation is shown in Figures 1-2.

\section{Dyon Helicity and a new SUSY.}

Let us now 
focus our attention to a single $j = const.$ sector. The spectra on FIGs.
1-2. are reminiscent of those of a SUSY system except for non-zero 
ground-state energy: (3.19) with $n = 0$ i. e. $p = p_0 = L_0 +
1 = j - 1/2$ yields indeed 
\begin{equation}
E_0^{(j)} =  q^2 \Big( 1 - ({q \over \vert \gamma \vert})^2 \Big) =
q^2 \Big( 1 - ({q \over
j + 1/2})^2 \Big), 
\end{equation}
since $\vert \gamma \vert$ = $j+1/2$. 
Let us  subtract therefore
the ground-state energy $E_0^{(j)}$ from the Hamiltonian and consider rather 
\begin{equation}
\left(\begin{array}{ll}
K_1 &\\  &K_0\end{array}\right)
= 
\D^2 - E_0^{(j)}
= \left(\begin{array}{ll}H_1-q^2 + {q^4 \over \gamma^2} &\\ 
 &H_0 - q^2 + {q^4 \over \gamma^2}
 \end{array}\right).
\end{equation}

Now we show that the new Hamiltonian (4.2) is indeed supersymmetric:
following Biedenharn [7-9], let us define the 'Dyon Helicity' operator
${\cal R}$ as follows: set first 
\begin{equation}
R = w \Big[ \Big(\partial_r+ displaystyle{1\over r}-\displaystyle{y \over r}\Big) y + q^2 \Big],
\end{equation}
so that
\begin{equation}
\begin{array}{ll}
R &= 
\Big[ \Big(- (\partial_r +\displaystyle{1\over r}) - 
\displaystyle{x\over r}\Big)x + q^2\Big]
w,
\\[6pt] 
R^{\dagger} &= 
w \Big[ \Big(\partial_r+\displaystyle{1\over r}-\displaystyle{x\over r} \Big)x+q^2\Big]
=\Big[\Big(-(\partial_r +\displaystyle{1\over r})-\displaystyle{y \over r} \Big)y+q^2\Big] w.
\end{array} 
\end{equation} 

Since 
$xR = - Ry$ and $yR^{\dagger} = - R^{\dagger} x$,
it is easy to verify that 
\begin{equation}
R R^{\dagger} = (2H_0 - q^2) x^2 + q^4 \quad \hbox{and} \quad
 R^{\dagger} R = (2H_1 - q^2) y^2 + q^4 .
\end{equation}

Generalizing Bienharn's approach [7-9], let us now introduce the 'Dyon
Helicity' or 'Biedenharn-Johnson-Lippmann' [11] operator as
\begin{equation}
{\cal R} = \left(\begin{array}{ll} &R^{\dagger} \\ R &\end{array}\right).
\end{equation}
${\cal R}$ satisfies the (anti)commutation relations, 
\begin{equation}
\big[{\cal R}, {\bf J} \big] = 0,
\qquad
\big\lbrace{\cal R}, \Gamma \big\rbrace = 0,
\qquad
\big\lbrace{\cal R}, \rho_3 \big\rbrace = 0,
\end{equation}
and
\begin{equation}
{\cal R}^2 = 
\left(\begin{array}{ll}R^{\dagger}R &\\  &R R^{\dagger}
\end{array}\right) = 
(\D^2 - q^2) \Gamma^2 + q^4.
\end{equation}
Our Dyon Helicity operator ${\cal R}$ preserves thus the total angular
momentum $j$, changes the sign of $\Gamma$, and interchanges the chiral
eigensectors. 

It follows now from (4.6), that, for each $j =
const.$ sector, we get hence a {\it new supersymmetric system} 
with supercharges
\begin{equation}
{\cal S} = \left(\begin{array}{ll} &S^{\dagger} \\ S &
\end{array}\right) =
\left(\begin{array}{ll}
\displaystyle 
&w \Big[ \partial_r + {1\over r} - {x \over r}  + {q^2 \over x}
\Big] \\  w \Big[ \partial_r + {1\over r} - {y \over r} 
 + {q^2 \over y}
\Big] &\end{array}\right). 
\end{equation}
In fact, 
\begin{equation}
\left(\begin{array}{ll}
K_1 &\\  &K_0 
\end{array}\right)
  =  {\cal S}^2,
\end{equation}
and (4.7) implies the analogous relations
\begin{equation}
\big[{\cal S}, {\bf J} \big] = 0, \qquad
\big\lbrace{\cal S}, \Gamma \big\rbrace = 0,
\qquad
\big\lbrace{\cal S}, \rho_3 \big\rbrace = 0.
\end{equation}

The identities
$$
{\sigmab\cdot\pib} = ({\sigmab}\cdot{\hat {\bf r}}) ({\hat {\bf r}}
\cdot{\pib}) 
+ ({\sigmab }\times {\hat {\bf r}}) ({\hat {\bf r}} \times {\bf
\pib}) 
= ({\sigmab .} {\hat {\bf r}}) ({\hat {\bf r}}.{\bf \pib}) 
+ i ({\sigmab .} {\hat {\bf r}}) ({{\sigmab .} {\bf \ell} \over r})
$$
imply that
\begin{equation}
i\sigmab\cdot\pib = w (\partial_r + {1 \over r} - {z \over r}),
\end{equation}
and thus $S$ is indeed (1.11).

For positive-energy states we can thus define the new
the unitary transformations
\begin{equation}
V = S \,{1 \over\sqrt{S^{\dagger}S}},
\qquad 
V^{-1} = V^{\dagger} = \ {1 \over \sqrt{S^{\dagger}S}} \, S^{\dagger}, 
\end{equation}
which intertwine the
eigensectors of the '$\Gamma$ - fermionic operator' 
\begin{equation}
{\rm sign}\,\Gamma = {\Gamma \over  \mid \gamma \mid}\ \ , 
\end{equation}
$V^{\dagger} K_1 V = K_0$.
Plainly, $x$ (resp. $y$) is conserved for the $K_0$  (resp.
$K_1$) dynamics,  $[x,K_0] = [x,H_0] = 0$, 
and $[y, K_1] = [y, H_1] = 0$.
The new SUSY transformations interchange  $x$ and $-y$,
\begin{equation}
V^{\dagger} y V = -x \qquad V x V^{\dagger} = - y.
\end{equation}

For $j \geq q +1/2$, each $j=const.$ sector has $2(2j+1)$
ground-states: while the $2j+1$ counts the angular eigenfunctions
$\varphi$ and $\Phi$
constructed in Section 3, the $2$ comes from the two single lower-lying dots in
each $j$-column in FIG.1-2. These states have wave number 
$k_0 = iq^2 /\vert \gamma \vert = iq^2 / (j + 1/2)$,
and the wave function is
\begin{equation}
u_0^{(j)} = r^{\vert \gamma \vert-1} e^{-{q^2 \over \vert \gamma \vert} r} 
= r^{j-1/2} e^{-{q^2 \over j+1/2} r}.
\end{equation}
(times the corresponding angular eigenfunctions $\varphi$ and $\Phi$), 
which
clearly generalizes (3.21) to $j \geq q + 1/2$. For these values of the angular
momentum, the Atiyah-Singer index (defined as the difference of the number of
solutions of ${\cal S} \phi = 0$ in the different supersectors) is 0. For the
lowest angular-mementum value $j = q -1/2$ the ground state has vanishing
energy and thus $S = Q$, so we recover those ground states (3.21) in the
'upper' sector with Atiyah-Singer index $2q$.

It is amusing to see how these properties are verified in the explicit angular
basis described in the previuos Section: if $\Psi$ is an
$E$-eigenstate of $-\D^2$ in the $\gamma$-sector, 
$\Gamma  \Psi = \gamma
\Psi$, then $\Gamma ({\cal S} \Psi) = - {\cal S} (\Gamma \Psi) = - \gamma
({\cal S} \Psi)$, 
so that $V \Psi$ is a state with the same energy in the $(-\gamma)$
sector. More precisely, let us consider a $\gamma$-eigenstate $\Psi_{\gamma}$ =
$u_{\pm} \Phi_{\pm}$. Since $w\Phi_{\pm} = \varphi_{\mp}$,
\begin{equation}
{\cal S} \Psi_{\gamma} = \Big[ ( \partial_r + {1 + \gamma \over r}
-  {q^2 \over \gamma}) u_{\pm} \Big] \, \varphi_{\mp} ,
\end{equation}
the action of the Dyon Helicity operator decomposes into radial and
angular action, and the angular part ($w$) just switches over the angular
eigenfunctions. But using the recurrence relations
 of the hypergeometric functions one proves [7], that 
\begin{equation}
\Big[ \partial_r + {1 + \gamma \over r} - {q^2\over\gamma} \Big] u_{\pm} 
= \sqrt{k^2 + {q^4\over \gamma^2}} \,\ u_{\mp},
\end{equation} 
so that the two factors combine into 
\begin{equation}
V(\Psi_{\pm\mid\gamma\mid}) = \psi_{\mp\mid\gamma\mid} \qquad \hbox{and}
\qquad V^{\dagger}(\psi_{\pm\mid\gamma\mid}) =
\Psi_{\mp\mid\gamma\mid}, 
\end{equation}
i.e. the new SUSY just intertwines the wavefunctions.

It is interesting to note, that
{\it both} ground states $\Psi _0$ and $\psi_0$ satisfy the
first-order relations
\begin{equation}
\begin{array}{llll}
{\cal S} \Psi _0 = 0 
&\hbox{and}
&{\cal S}^{\dagger} \psi _0 = 0 \qquad &\hbox{for }\, j \geq q+1/2
 \\
& & & 
\\
{\cal S} \Psi _0 = 0 & & &\hbox{for }\, j = q-1/2, 
\end{array}
\end{equation}
where we dropped the upper index $j$ for simplicity. Indeed, for the 'upper'
(i.e. $\rho_3 = +1$) sector the ground state corresponds to $y = - \gamma =
\vert \gamma \vert$ and in in the 'lower' i.e. (i.e. $\rho_3 = -1$) sector
$x = - \gamma = \vert \gamma \vert$. Therefore the radial parts of ${\cal S}$
and ${\cal S}^{\dagger}$ are, up to an overall sign, identical. But the
ground-state wave number is $k_0^2 = -q^4/\vert \gamma \vert$ and thus our
statement follows from (4.16) and (4.17).

\section{The S - matrix from SUSY.}

For large r the 1/r terms can be dropped in the Dyon Helicity operator, 
\begin{equation}
R \rightarrow  R_{scatt} = (-i p_r x - q^2 ) w
\quad \hbox{and} \quad 
R^{\dagger} \rightarrow R^{\dagger}_{scatt} = (-i p_r y - q^2) w.
\end{equation}
Their actions on the scattering states are therefore
\begin{equation}
R_{scatt}^{\dagger} \Psi^{\epsilon}_{\pm} = (-i\epsilon k \gamma - q^2 ) 
\psi^{\epsilon}_{\mp} \quad \hbox{and} \quad 
R_{scatt} \psi^{\epsilon}_{\pm} = (-i\epsilon k \gamma - q^2 ) 
\Psi^{\epsilon}_{\mp},
\end{equation}
where $\epsilon = \pm 1$ for in/out.

The eigenstates $\Psi_{\pm}$ and $\psi_{\pm}$ can
asymptotically be expanded as
\begin{equation}
\Psi_{\pm} \sim A^{in}_{\pm} \Psi_{\pm}^{in} + 
A^{out}_{\pm} \Psi_{\pm}^{out}, \qquad \,
\psi_{\pm} \sim a^{in}_{\pm} \psi_{\pm}^{in} + 
a^{out}_{\pm} \psi_{\pm}^{out}
\end{equation}

Chose $A = a$ as in the Coulomb case [9]. Acting on $\Psi_{\pm}$ in the r.h.s.
of (5.3) by $R$ produces a state in the opposite sector with shifted index,   
which is expanded as
\begin{equation}
R(\Psi_{\pm}) \sim \sqrt{k^2 \gamma^2 + q^4}\ (a^{in}_{\mp} \psi_{\mp}^{in} + 
a^{out}_{\mp} \psi_{\mp}^{out})
\end{equation}
Acting on the l.h.s. of (5.3) gives in turn, by (5.2),
\begin{equation}
(-i\epsilon k\gamma - q^2) \ (A^{in}_{\pm} \psi^{in}_{\mp} + 
A^{out}_{\pm} \psi^{out}_{\mp})
\end{equation}
Equating (5.4) and (5.5) yields the relation
\begin{equation}
{A^{\epsilon}_+ \over A^{\epsilon}_-} = \sqrt{{\epsilon \gamma +iq^2/k 
\over \epsilon \gamma +iq^2/k}} .
\end{equation}
Now remember that the $\pm$ signs actually mean $\gamma$ and $\gamma - 1$, 
$A_+ = A_{\gamma}$ and $A_- = A_{\gamma - 1}$ so
(5.6) can be viewed as recursion relations whose solutions are
\begin{equation}
A_{\gamma}^{\epsilon} = 
(i\epsilon)^{\gamma} \sqrt{{\Gamma (\gamma + i\epsilon q^2/k) \over
\Gamma (\gamma -i\epsilon q^2/k)}},
\end{equation}
yielding the phase shift
\begin{equation}
{A_+ \over A_-} = (-1)^{\gamma} {\Gamma (\gamma +iq^2/k) \over
\Gamma (\gamma -iq^2/k)} \times \hbox{(a $\gamma$ - independent constant).}
\end{equation}
The S-matrix is hence
\begin{equation}
S_{\gamma} = exp \ (2i\delta_{\gamma}) \ C(k), \quad \hbox{where} \quad
\delta_{\gamma} = \arg \ (\Gamma(\gamma + 1 + iq^2/k)).
\end{equation}
The poles of the $\Gamma$-function yield once more the positive bound-state
spectrum (3.19). The result is consistent with the one obtained in Ref. 2 using
the dynamical symmetry.

\section {Discussion.}

Some aspects of the dyon problem would require further study. 
One of these concerns the {\it dynamical symmetries}: Both
Hamiltonians $H_0$ and $H_1$ 
have conserved Runge-Lenz vectors
as well as extra conserved spin-type vectors, which
combine with the angular momentum into an $o(4) \oplus o(3)$ dynamical
symmetry [1, 4]. The spectrum represented on FIGs. 1-2. appears to be
consistent with the extension to $o(4,2)$ [20]. The technique
of Barut and Bornzin [21] allows in fact to build an $o(2,1)$ which should
furthermore
 combine with the $o(4)$, to provide a spin-dependent realization of $o(4,2)$,
different from the classic one [22].
A related question is to clarify the
relation between the Runge-Lenz vector, the spin vectors and the 
Dyon Helicity operator.

A second remark concern the individual $j=const.$ and $\rho_3 =$ fixed sectors 
in FIGs. 1-2.: not only do we get the same patterns in the upper and lower sectors,
but, within each sector, the two equations are actually the {\it same} up to a
shift $j \to j+1$. This seems to indicate a {\it shape invariance} [23].

\vspace{3mm}
\noindent
{\bf Acknowledgements.} We would like to thank Professor Z. Horv\'ath for
calling our attention to Ref. 13 and Dr. N. Backhouse for providing us with 
Ref. 8, which lead us to a breakthrough. One of us (P. A. H.) is indebted to
the University of Liverpool and to the Dublin Institute for Advanced Studies
for hospitality and partial financial support, and to Professors L. C.
Biedenharn and L. O'Raifeartaigh for enlightening discussions.


\end{document}